\documentclass{artikel3}
\usepackage[T1]{fontenc}
\usepackage[portuges,english]{babel}
\usepackage{amssymb,latexsym,amsmath,color,setspace}
\usepackage[colorlinks,linkcolor=blue,urlcolor=blue,citecolor=black,
plainpages=false,pdfpagelabels,breaklinks]{hyperref}
\usepackage{stmaryrd}

\newtheorem{thm}{Theorem}[section]

\newtheorem{dfn}{Definition}[section]

\newcommand{\tbf}{\textbf}
\newcommand{\ita}{\textit}
\newcommand{\mcal}{\mathcal}

\newcommand{\qst}{$\mathfrak{Q}$}
\newcommand{\igual}{:=}

\title{Presenting Nonreflexive Quantum Mechanics: Formalism and Metaphysics}

\author{{D\'ecio Krause}\thanks{Partially supported by CNPq.} \and  {Jonas R. B. Arenhart}}
\date{\small Research Group in Logic and Foundations of Science\\ Department of Philosophy\\ Federal University of Santa Catarina \\ (Forthcoming in \ita{Cadernos de História e Filosofia da Ciência}, Campinas - SP, Brazil) \\ October 2015}

\begin{document}
\maketitle

\begin{abstract} Nonreflexive quantum mechanics is a formulation of quantum theory based on a non-classical logic termed \ita{nonreflexive logic} (a.k.a. `non-reflexive'). In these logics, the standard notion of identity, as encapsulated in classical logic and set theories, does not hold in full. The basic aim of this kind of approach to quantum mechanics is to take seriously the claim made by some authors according to whom quantum particles are \ita{non-individuals} in some sense, and also to take into account the fact that they may be absolutely indistinguishable (or indiscernible). The nonreflexive formulation of quantum theory assumes these features of the objects already at the level of the underlying logic, so that no use is required of symmetrization postulates or other mathematical devices that serve to pretend that the objects are indiscernible (when they are not: all objects that obey classical logic are \ita{individuals} in a sense). Here, we present the ideas of the development of nonreflexive quantum mechanics and discuss some philosophical (mainly metaphysical) motivations and consequences of it.
\end{abstract}

\section{Introduction}

Indiscernibility of physical systems (or physical objects) appear already in classical physics. For instance, in 1808 John Dalton, the founder of modern atomism, claimed that there would be no differences among chemical elements of a same kind: ``we may conclude that the ultimate particles of all homogeneous bodies are perfectly alike in weight, figure, etc.'' \cite[p.142-3]{dal08}. In 1900, Max Planck, still reasoning within the scope of classical physics, made the fundamental move to quantum mechanics by assuming that in distributing an integral number $N$ of energy elements over $C$ states, the indiscernibility of the elements should be taken into account. Planck's formula
$$Z = \frac{(N+C-1)!}{N! (C-1)!}$$
\noindent shows that the division by $N!$ makes the trick. Some years before, the so-called Gibbs paradox  appeared in classical statistical mechanics showing that indiscernible particles are not only part of the physics, but also of the philosophical problem concerning the proper understanding of how an appropriate metaphysics for physical theories would look like. Indeed, should we take the natural approach suggested by the Sackur-Tetrode formula for the entropy of an ideal gas and accept that nature made particles indistinguishable and that there is no more talking about the topic? Or should we adopt a minimalist approach and concede that indistinguishability may help us in drawing the line between classical and quantum particles (see French and Krause \cite[chap.2]{frekra06})? The first route would blur the difference between classical and quantum particles, both being absolutely indistinguishable in some cases. In the second case, we may argue that classical particles are different from their quantum counterparts for they may be considered to be individuals and distinguishable in some sense, while the matter is not so easily settled for quantum particles, which, due to their indiscernibility, may be taken as some kind of \ita{non-individuals}.

The word `non-individuals' has an historical use. Several forerunners of quantum theory have referred to quantum objects this way, mainly in the sense that quantum particles ``are not individuals'', as suggested by Schrödinger \cite[p.206]{sch57}. But the term `non-individual' may confuse the reader in suggesting that these entities would be something different from that which we can refer to or speak about. In fact, we don't know in precise terms what they are. Each theory gives its own characteristics to these entities (B. Falkenburg traces the `metamorphoses' of the term `particle' from classical physics to the most advanced physical theory of today \cite[Chap.6]{fal07}). Being non-individuals does not prevent quantum particles from being able to be isolated by some time, but that they cannot be regarded as individuals \ita{in the standard sense of the word}, as something that have an identity and can be identified in other times even if mixed with others of similar species. As it is well known, this does not happen with quantum particles. It is in this sense that we say that all protons are absolutely indiscernible, so are all electrons, all neutrons, and do on. Hence, all atoms of the same chemical isotope, made from protons, neutrons and electrons, are also indiscernible. Without this hypothesis, chemistry does not work, and quantum physics does not work either.

We shall take the second approach mentioned earlier and concede that there is a significant difference between quantum and classical particles. In classical mechanics we can at least in principle follow well-defined trajectories for each particle and by adding a reasonable assumption concerning their impenetrability we may grant that some form of spatio-temporal principle of individuality holds. When we turn to quantum mechanics, on the other hand, things get much more complicated; in fact, for quantum particles there is no trajectory in the classical sense, unless a hidden-variables approach is followed, which has its own shortcomings; the impenetrability assumption is also unavailable, for it may happen that particles are in entangled states. More than that, quantum mechanics encapsulates indistinguishability in the statistical treatment of aggregates of particles through the famous Bose-Einstein and Fermi-Dirac statistics. These are seen as grounding the fact that in some cases nothing whatsoever may distinguish quantum particles, for as is famously remarked, permutations of particles do not count in this statistics as giving rise to different states.

One of the difficulties with this fact --- quantum indistinguishability as obtained through the statistics of the theory --- concerns precisely the relationship between the experimental fact that is being expressed and its mathematical formulation. Experimentally, it is found that there are situations in which one cannot distinguish between two particles of the same kind, as the theory predicts; mathematically, though, since the formalism of the theory is constructed within classical set theory using a part of functional analysis, the indistinguishability must be brought into the theory through the assumption of some symmetry conditions that grant us the desired effect. In a standard set theory, let us emphasize, whenever we have \ita{two} objects, they are necessarily discernible: they are \ita{distinct}. Now, from a philosophical point of view the assumption of these symmetry conditions seems to be at odds with the idea of genuinely indistinguishable entities, for it seems to be a mechanism to mask the fact that the entities dealt with are not really indistinguishable, since they belong to classical logic and set theory (see the discussions in Krause and Arenhart \cite{kraare15}).

The difficulty involving standard mathematics and logic and their relation to `genuine' indistinguishability can be expressed in a nutshell as follows: on the one hand, orthodox quantum mechanics says that there are situations according to which we cannot discern the quanta; on the other hand, within classical mathematics, we represent this fact using symmetry postulates,  whose aim is to mask the fact that in standard logic and mathematics one can always know that wherever there are two items, they are discernible by purely logical means inside set theory. We may even opt for a realistic interpretation according to which, even conceding indistinguishability, the particles \ita{remain there} with their hidden properties, and even in this case there is no way to tell them apart. So, concerning this situation, three options seem to be available: (i) either the problem concerns the relationship between the entities dealt with by QM and standard mathematics, so that those entities may not sit comfortably together with assumptions made in classical mathematics, or (ii) there is some kind of hidden variable which may confer individuality to quantum entities, but, being hidden, is not taken into account by the standard formalism, or (iii) the quanta must be treated mathematically as certain entities which are discernible but their discernibility cannot be expressed in qualitative terms by the vocabulary of the theory. Perhaps the last two may be seen as just one option, but nothing in principle prevents us from distinguishing them. In fact, (ii) allows that in principle we could extend the formalism to accommodate the hidden variables accounting for the individuality, something (iii) in this reading does not allow for; in this last case the individuality is there but the theory does not allow us to express how it is constituted.

All these options present their own challenges; after briefly discussing items (ii) and (iii) in sections \ref{ind} and \ref{sta}, we turn to the first one. This is perhaps the place to say that we are not claiming that QM violates standard mathematics and logic; our point is that it seems that this theory can be formulated in different and non equivalent ways, most of them within these `classical' frameworks. In saying that, we are not assuming that there is just one QM, we take it as a fact that there are different formulations of different theories which are empirically equivalent. What we intent to do is to explore an alternative approach and look for the (at least conceptual) gains we may have. As we shall see in section \ref{qsp}, there is an alternative formulation/construction of quantum mechanics that incorporates non-individuality right from the start and makes use of this fact in order to develop the formalism of the theory. It is our goal here to render this formulation more easily accessible and clear. Once that is done, the philosophical view that quantum particles are not individuals (along with some simple logical theory of quasi-set) is seen as giving rise to fruitful developments of the theory itself, as an instance of a kind of reflexive equilibrium procedure: the standard mathematical formulation and experiments suggest that quantum particles are not individuals; however, that mathematical apparatus does not reflect that non-individuality, so, we use that fact in order to develop a mathematical formalism that takes non-individuality into account; this formalism, now, gives rise to another version of quantum mechanics. We discuss this view in section \ref{dis}.

\section{Indistinguishability as a quantum phenomenon}\label{ind}

To better understand the reasons why quantum particles are considered to be absolutely indistinguishable and how that absolute individuality is supposed to lead to their \ita{non-individuality}, we may start by comparing quantum and classical statistics. This is the standard \ita{trope} on this issue, and we believe it is a good strategy to be followed, because since we have no access to the particles themselves, it seems that their collective behavior may give us some clues as to their identity and (in)distinguishability (see French and Krause \cite[chap.4]{frekra06}).

The first point to be noticed is that there is some sense in which both classical and quantum particles are taken to be indistinguishable: when they share all their intrinsic (or state independent) properties. Sometimes it is common to call particles sharing all their intrinsic properties  `identical particles'. However, this is just an abuse of language to say that the particles are of the same kind. The similarities among quantum and classical stop there, however. The main feature behind the claim that classical particles are individuals after all and that they are not really indistinguishable in the same sense that quantum particles are said to be concerns their collective behavior; the way that classical statistical mechanics describes aggregates of classical particles differs significantly from the way quantum mechanics describes aggregates of quantum particles.

Classical particles behave collectively according to Maxwell-Boltzmann statistics (MB). Let us illustrate it by considering the particular case in which there are two particles, labeled $1$ and $2$, to be distributed in two different states, $A$ and $B$. Then, according to MB statistics, we can have the following cases:
\bigskip
\begin{enumerate}
\item A(1) and A(2);
\smallskip
\item B(1) and B(2);
\smallskip
\item A(1) and B(2);
\smallskip
\item A(2) and B(1).
\end{enumerate}
\bigskip
Assuming the equiprobability hypothesis --- the claim that none of these cases has more chances to occur than the others (a claim not without its difficulties, see French and Krause \cite[chap. 2]{frekra06}) --- implies that all these possibilities are assigned the same weight, that is, $\frac{1}{4}$, and despite being considered as indistinguishable in the sense discussed above, permutations of particles, as it happens in the cases 3 and 4, are counted as giving rise to distinct possibilities. Here lies the key where the alleged difference between classical and quantum particles concerning identity and individuality matters.

The fact that permutations should bee seen as giving rise to different situations is generally taken to witness for their individuality. In fact, even though the particles share all their intrinsic properties, this simple fact seems to speak in favor of some individuality being manifested by the particles as they are permuted. Their individuality accounts for the difference in cases 3 and 4. Another way to put it is in terms involving state descriptions as describing a possible world: a permutation of particles in case 3 leads to a distinct possible world, one in which it is particle 2 that is in A and 1 in B. What accounts for the differences in the two worlds? Not the properties of the particles, assuming them indiscernible. So, it must be the case that they have an individuality intrinsic to them that is responsible for this possibility. That individuality can be understood in terms of some underlying substratum, a \ita{haecceity}, or even more simply in terms of the trajectories (spatio-temporal location) with the impenetrability assumption of which we have already spoken about earlier (see, for instance, Ladyman \cite{lad07} for a discussion in those terms).

In quantum mechanics, on the other hand, nothing like this happens. To keep with the case of two particles labeled once again $1$ and $2$ distributed in states $A$ and $B$, we have two cases, one for bosons and other for fermions. For the bosons, the following possibilities obtain:
\bigskip
\begin{enumerate}
\item A(1) and A(2);
\medskip
\item B(1) and B(2);
\medskip
\item A(*) and B(*).
\end{enumerate}
\bigskip
Again assuming equiprobability for each of these cases, each of them will have the same probability of $\frac{1}{3}$ to occur. Notice that in case 3 we have employed $*$ instead of the particles labels; that accounts for the distinctive feature of quantum statistics: the fact that permutations do not give rise to a distinct situation. The $*$ then means \ita{one} particle in $A$ and \ita{one} in $B$. The same idea can be used for Fermi-Dirac statistics, the one employed for fermions:
\bigskip
\begin{enumerate}
\item A(*) and B(*).
\end{enumerate}
\bigskip
For fermions, as for bosons, permutations change nothing, but differently from bosons, the fermions obey the famous Pauli Exclusion Principle, which roughly speaking grants that no two fermions can occupy the same state; thus there is only one option in this particular case. The great novelty of quantum statistics comes from the fact that permutations do not give rise to distinct states. In terms of the description involving possible worlds above, a permutation does not give rise to a distinct world. Contrarily to the classical case, there is nothing in the particles to ground the fact that the situation before the permutation is distinct from the situation after the permutation. Thus, they are the same situation. As a consequence, there is no intrinsic individuality principle doing that job.

Obviously, if that kind of reasoning is going to ground the claim that quantum particles have no individuality and are really indistinguishable, as we wish it to do, then it must be recognized that our language is inadequate to express that fact, for we began by labeling the particles, which amounts to individuate and identify them (something we cannot do for absolutely indistinguishable items), and then we were forced to `forget' the labels and place an $*$ in their place to neutralize the fact that if a label was allowed in the counting we would have intuitively the same classical statistics given by Maxwell-Boltzmann. Now, this is precisely the conundrum between the ``nature'' of quantum particles and the language we use to talk about them. To deal more properly with this topic, let us see how the formalism of quantum mechanics deals with the statistics. This will help us to appreciate the relevance of the alternative formulation of quantum mechanics we introduce later.

\section{The standard way to deal with indistinguishable objects in QM}\label{sta}

We would like to emphasize that in standard approaches to the formalism of quantum mechanics, given inside some classical set theory, quanta are individuals (as entities represented either by sets or by ur-elements) whose individuality is veiled by symmetry principles that are added by hand. In fact, the standard formalism of quantum mechanics masks the individuality of the particles through the use of a mathematical trick: one begins with labelled individuals (particles 1 and 2, say) and by convenient manipulation, a clever postulate mimics the behavior of indiscernible non-individuals, say by means of symmetric and anti-symmetric vectors/functions. The price to be paid for the use of such a \ita{man{\oe}uvre} is mainly philosophical, since physics works for all practical purposes: one wants to be committed with an ontology of non-individuals, but in the end, all we are left with are the same old individuals we started with, clothed as non-individuals by the symmetry postulates asserting the invariance of their permutability. Let us see precisely how it is done, for when we know what we wish to avoid, it is then easier to fully appreciate what is gained by the use of the \qst-spaces introduced below. Also, this way of approaching quantum mechanics, through the use of labels (we don't know how to proceed without them within such standard mathematical frameworks) is one of the signs of individuality.

Consider the quantum statistics mentioned above. To each  particle $1$ and $2$ we associate a Hilbert space $\mcal{H}_{1}$ and $\mcal{H}_{2}$ respectively. If the particles are indiscernible, these spaces are in fact the same for both. For the system composed by both particles we associate the tensor product $\mcal{H}_{1} \otimes \mcal{H}_{2}$. As is usual, we write a typical vector of this space by $|\psi_1\rangle \otimes |\psi_2\rangle$, or simply $|\psi_1\rangle |\psi_2\rangle$ for short. Then, the possibilities available for the particles $1$ and $2$ in states $A$ and $B$ are stated as the following:
\bigskip
\begin{enumerate}
\item $|\psi_{1}^A\rangle |\psi_{2}^A\rangle$
\medskip
\item $|\psi_{1}^B\rangle |\psi_{2}^B\rangle$
\medskip
\item $\frac{1}{\sqrt{2}}(|\psi_{1}^A\rangle |\psi_{2}^B\rangle \pm |\psi_{2}^A\rangle |\psi_{1}^B\rangle)$.
\end{enumerate}
\bigskip
We are here simplifying the notation: the third case splits in two, according to whether we deal with bosons or fermions; for bosons, we have cases 1-3 with a plus sign in the last one (symmetric vector), for fermions we only have the third possibility, now with the minus sign (anti-symmetric vector). What really matters is that to obtain the correct statistics one must employ symmetric and anti-symmetric vectors. This is the first part of the trick: label the particles and write the appropriate states, that is, symmetrical for bosons and anti-symmetrical for fermions. This means in particular that the asymmetrical states $|\psi_{1}^A\rangle |\psi_{2}^B\rangle$ and $|\psi_{2}^A\rangle |\psi_{1}^B\rangle$, like the ones employed in classical statistics, are not taken in consideration.\footnote{They constitute \ita{surplus formal structures}, according to a terminology introduced by M. Redhead, that is, elements of the formalism corresponding to nothing in the real world (see \cite[p.25]{tel95}).} Notice that the use of these states allows us to distinguish between the particles: we are able to attribute specific states to each one of them, although indiscernible (this is one way to substantiate the claim that quantum particles are indiscernible individuals; see French and Krause \cite[chap.4]{frekra06} for the details).

But how can we make sure that only the states with the appropriate symmetry types are available? That is the second part of the trick, and it consists in postulating that only the appropriately symmetrized states obtain. The famous \ita{Indistinguishability Postulate} (IP) does just this job:

\begin{equation}\label{IP}
\langle \psi_{12} | \hat{O} | \psi_{12} \rangle =  \langle P\psi_{21} | \hat{O} | P\psi_{21} \rangle.
\end{equation}

What this principle says, roughly speaking, is that the result of a measurement before and after a permutation of the labels of particles results in the same quantity, so, permuting the labels amounts to no physical difference at all. In fact, a more technical reading of IP says that only observables compatible with the permutation operators $P$ (the ones whose purpose is to permute the labels given to the particles) are allowed. This amounts to the following relation:

\begin{equation}
[\hat{O}P,P\hat{O}] = \hat{O}P - P\hat{O} = 0
\end{equation}

As one can check, only symmetrical and anti-symmetrical states satisfy these conditions. There is a whole discussion on what are the grounds to postulate IP and to accept it, and the different readings of this principle that may be compatible with it; in one of them, for example, which makes it weaker, the asymmetric states are not banned, but remain simply inaccessible to the particles (as we mentioned previously). In this last case, the particles may be regarded as individuals of some kind, but we shall not enter into these discussions here (but see French and Krause \cite[chap.4]{frekra06} for further discussions). We just remark that this possibility shows that quantum particles \ita{can} be considered as individuals too! (But, as we are trying to show, this is not the better way to see them).\footnote{In fact, quantum physics also \ita{can} (and \ita{is}) be developed within a standard set theory which comprises individuals; as we have already remarked, physics works fine, but the philosophical problems turn to light.}

So, this is in short how the trick is done in the standard formalism of quantum mechanics: to grant that permutations of particles do not give rise to different states, that is, to grant that the particles are absolutely indiscernible, one begins by labeling them in the first place, and then, by using only certain symmetrized vectors  and by imposing a condition that only these vectors represent physical states, we achieve the desired result.

Each of these steps has its own difficulties for the philosopher interested in a commitment with `real' indistinguishable entities, in particular for a defense of a metaphysics of nonindividuals. Really, to begin with, by assuming that there is some sense in labeling particles, we may consider them to have well defined identity conditions, for  one of them was labeled $1$ and \emph{the other} was labeled $2$. This idea runs counter to the whole enterprise being pursued, that is, that the particles must be \ita{really} indiscernible. As they are generally understood, labels seem to demand a semantic theory committed with individuals and identity.

Here it would be interesting to comment that even the Fock space formalism (which will be deal with below) is not completely free from initially attached labels, as Redhead and Teller have claimed (see \cite{tel95}). In fact, in their usual construction, they begin with labeled Hilbert spaces (spaces $\mcal{H}_{1}$ and $\mcal{H}_{2}$, as we have been discussing); so, even if in the final presentation of a Fock space the labels are not mentioned, we need labels in Hilbert spaces for the usual constructions of the Fock spaces. The trick is not, after all, dispensed with! This criticism justifies our approach in the next section to construct Fock spaces inside a nonreflexive system of logic in which none of the mentioned tricks is required. We begin with indiscernible objects and build from there.

Note: --- in fact, for $n$ indistinguishable particles, one takes the same Hilbert space $n$ times. But this does not eliminate the trick, for we still work with vectors of the form $| n_1, n_2, \ldots, n_k\rangle$, representing a state with $n_i$ quantum particles with a certain eigenvalue in a certain state (which would attest only to the quantity of them but not their individuality). In pursuing the details, however, we need to go back to sets and individuals. There is no scape within standard mathematics!

\section{The \qst-spaces}\label{qsp}

In order to avoid the (even hidden) assumption that we need to start with individuals, we introduce the notion of \qst-spaces.
We begin with an outline of quasi-set theory, the framework within which we develop our nonreflexive quantum mechanics.

\subsection{Basic quasi-set theory \qst}

Quasi-set theory is a ZFU-like first-order set theory comprising two kinds of atoms:\footnote{We work with atoms because it seems to be more intuitive to regard physical objects not as sets. As is well known, atoms are entities that can belong to sets, but which have no elements (in the sense of the membership relation), yet they could be \ita{composed} of other entities in a mereological sense. The problem is that a \ita{quantum mereology} was not developed yet -- regarding mereology, see \cite{sim87}; concerning some problems related to a quantum mereology, see \cite{kra11}.} the M-atoms, which behave like ZFU atoms and are denoted in the language with the help of a unary predicate $M$, and the m-atoms, which satisfy the unary predicate $m$. For the last kind of atoms identity is not defined, for according to our intuitive interpretation of the theory, we attribute to them the role of quantum particles: they must be not only indistinguishable, but more than that, identity statements must not make sense for them. To grant that both goals are achieved, that is, that there is one kind of objects that can be related by an indistinguishability relation and be such that identity and difference statements do not make sense for them, we must in \qst\ depart from some of the original ZFU features. Let us check briefly how it can be done (for more details, see  \cite[chap.7]{frekra06}, \cite{frekra10}).

Note: --- before we begin with the theory, let us make another general remark. Why to advance the thesis that the notion of identity is senseless for the objects denoted by the m-atoms? The problem is that, if we assume that they obey the traditional theory of identity of classical logic (and set theory), once we have two of them, they are necessarily \ita{different}. Well, in this case, they must have some difference, given by a property (since we are, as usual, avoiding to suppose notions of substratum)! Well, either the physical theory presents this property or not. In the first case, the entities could not be indiscernible; in the second case, the difference would be hidden in the underlying logic, but even so \ita{it needs to be regarded as existing}. Then we have again two cases: either logic (or set theory) shows us the property or it does not. In both cases the entities turn to be individuals --- we remark that even in standard mathematics there are entities (real numbers, say) that are different but whose difference cannot be described by the theory (an example would be the two least elements of two disjoint sets of real numbers) according to a given well-order, which can be assumed to exist in the presence of the axiom of choice, but which cannot be described by a formula of the theory of sets).

Let us begin with the underlying logic of \qst: it is first-order logic \emph{without} identity. From a syntactical point of view, the postulates of this logic are the same as those of classical logic, but semantically they are understood differently. To understand this restriction one needs only to remember that classical semantics is developed inside a classical set theory like ZFC. If this kind of semantics were employed to the underlying logic of \qst, then we would be committing ourselves with precisely those features of classical logic we want to avoid, namely, that identity makes sense for every object and indistinguishability relations collapse in identity. Quantifier must be understood with some reservation: $\forall x \varphi(x)$ is to be not read as `for each object of the domain', which would entail their identification, but simply as `for all objects of the domain', meaning precisely this: for all objects of the domain. Similarly, $\exists x \varphi(x)$ means that \ita{some} object of the domain satisfies the formula $\varphi(x)$, and we need not to identify it, say by a name; for details, see  \cite{are14},\cite{arkra09}.

Besides atoms, the theory is designed to grant that we can build collections. By definition, a quasi-set (q-set for short) is something that is not an atom, and we distinguish them for they obey a primitive unary predicate $Q$; that is, $Q(x)$ indicates that $x$ is a q-set. Such collections are assumed to be formed in stages, as it happens in the classical cumulative hierarchy; for some of these collections, in some step m-atoms may be involved as member, but for others it may happen that no m-atom appear in any step of the formation of the q-set. To designate this last kind of q-set we introduce the primitive predicate $Z$, denoting what we call a \ita{classical q-set}, or simply a \ita{set}, and they turn to be copies of the sets of ZFU. Technically speaking, sets are collections with no m-atoms in their transitive closure. We call the M-atoms and sets \ita{classical things} of the theory. When restricted to the classical things, \qst\ allows us to develop all the standard mathematics that can be developed inside ZFU.

Identity must be introduced by definition. This fact allows us to manufacture an identity relation restricted to q-sets and M-atoms only, but not holding among m-atoms (to fit the intuitive interpretation we have attributed them). The definition is the following:

\begin{dfn}
$x = y \igual (Q(x) \wedge Q(y) \wedge \forall z (z \in x \leftrightarrow z \in y)) \vee (M(x) \wedge M(y) \wedge \forall z (x \in z \leftrightarrow y \in z))$.
\end{dfn}

According to the definition, equal q-sets are those that have the same members, and equal M-atoms are those that belong to the same q-sets. If there are m-atoms involved, since the notion of identity does not make sense to them, it may be the case that we are unable to effectively know whether two q-sets are identical, for we have no means to identify their elements. But things here work in the conditional: \ita{if} two q-sets have the same elements, \ita{then} they are identical and reciprocally.\footnote{We recall Bertrand Russell's definition of pure mathematics (and \qst\ is the pure counterpart of our quantum mechanics) as the class of propositions of the form `$p$ implies $q$', given in his \ita{The Principles of Mathematics}.}

One can easily prove from this definition that identity is reflexive, and the substitution law is postulated as an axiom; then, identity (where it holds) has both properties that characterize it in first-order systems. What about indistinguishability? We postulate that it has the properties of an equivalence relation (\ita{i.e.} it is reflexive, symmetric and transitive), but it is not compatible with the membership relation, that is, from $x\equiv y$ and $x \in z$ we cannot always derive that $y \in z$; this is done in order to grant that identity and indistinguishability do not collapse in the same concept. For the other properties $P$ of the language, though, we have that if $x \equiv y$ and $P(x)$, then $P(y)$. Also, for the classical things, identity and indistinguishability are equivalent.

The construction of q-sets proceeds in much the same way as in ZFU: by the iterated application of well-known set theoretical operations as power set, union, cartesian product, among others. One difference that is worth mentioning concerns the notion of unordered pair; in \qst\ we cannot proceed as in the usual set theories, postulating that for every $x$ and $y$ there is a collection of objects equal to either $x$ or $y$, and the reason is simple: for m-atoms identity is not defined, so that to restrict ourselves to the classical definition would amount to the impossibility of pairs of m-atoms, something undesirable in a theory of collections. To mend the situation, the theory encompasses an unordered pair axiom stating that for every two items $x$ and $y$ there is a q-set $z$ containing them as elements (as in the classical case, there may be other elements in $z$ as well). We then apply the separation axiom (which has its quasi-set theoretical version similar to the classical formulation) to a q-set $z$ and separate a `pair', in fact, which is the q-set of the elements of $z$ containing only those elements that are indistinguishable from either $x$ or $y$. This collection is denoted $[x, y]_{z}$; if $x \equiv y$ holds, then we write simply $[x]_{z}$ and call it the weak singleton of $x$. Notice that, intuitively speaking, the quantity of elements in weak pairs and weak singletons may be higher than 2 and 1 respectively, for $z$ may contain many elements indistinguishable from $x$ and/or $y$.

From unordered pairs we can then define a weak version of ordered pairs as follows, which stands for the q-set of the items indistinguishable from $x$ and from $y$ that belong to $z$:

\begin{dfn}[Weak ordered pair]\label{wp}
$\langle x, y \rangle_z \igual [[x]_z, [x,y]_z]_z$
\end{dfn}

Of course, any binary relation (or q-relation as we shall see) involving only m-atoms is symmetric for $\langle x, y \rangle_{z} = \langle y, x \rangle_{z}$.
Then, the definition of cartesian product follows as usual:

\begin{dfn}[Cartesian Product]
$z \times w \igual [\langle x, y \rangle_{z \cup w} : x \in z \wedge y \in w]$  \end{dfn}

If there are only m-atoms involved, then $z \times w = w \times z$ for all $z$ and $w$, as is easy to see. The theory has also a union axiom which works as in standard set theories; so, we shall use the symbols $\cup$ and $\cap$ taken then from granted from now on.

\begin{dfn}[Quasi-relation] A q-set $R$ is a binary quasi-relation between q-sets $z$ and $w$ if its elements are weak ordered pairs of the form $\langle x , y \rangle_{z \cup w}$, with $x \in z$ and $y \in w$. \end{dfn}

Now, for quasi-functions we cannot proceed as usual, for the classical concept of function uses the notion of identity. To generalize the usual concept, then, and allow that it applies also for q-sets of m-atoms, we require that a mapping does not attribute different \ita{kinds} of values to indistinguishable arguments. Thus, we have:

\begin{dfn}[Quasi-function]\label{qf} $f$ is a quasi-function among q-sets $A$ and $B$ if and only if $f$ is quasi-relation between $A$ and $B$ such that for every $u \in A$ there is a $v \in B$ such that if $\langle u, v\rangle \in f$ and $\langle w,z\rangle \in f$ and $u \equiv w$ then $v \equiv z$. \end{dfn}

What this definition grants us is that a quasi-function maps indistinguishable elements to indistinguishable elements. For classical objects, both quasi-relations and quasi-functions coincide with the classical definitions. With some restrictions, one can also define the concepts of injection, surjection and bijection, but we shall not present those definitions here (see \cite[chap. 7]{frekra06}).

Another important concept in \qst\ concerns the attribution of cardinality to q-sets. Since collections having m-atoms as elements cannot be well-ordered (for that concept also presupposes identity), one cannot proceed in the usual ways to grant that every q-set will have a cardinal number associated to it, for example, through the attribution of ordinals following the well-known von Neumann definition. The usual strategy has been to adopt a primitive concept of quasi-cardinal which generalizes the classical notion of cardinal: in the classical part of quasi-set theory one builds cardinals, as usual, and then, with the primitive quasi-function $qc$ one attributes cardinals to every q-set, making sure that classical sets will have as their quasi-cardinal (the cardinal attributed by $qc$) precisely the same cardinal attributed to them in the classical part of the theory.

The notion of quasi-cardinal also allows us to obtain in \qst\ another interesting concept: the \ita{strong singleton}. As we have seen, given any object $x$ whatever we may obtain the weak singleton of $x$, a q-set which in the case of $x$ being an m-atom may have quasi-cardinal greater than 1. The strong singleton of $x$ relative to a q-set $w$, denoted by $\llbracket x \rrbracket_w$ is simply a q-set containing an element that is indistinguishable from $x$ which belong to $w$ and whose quasi-cardinal is precisely 1. With the resources of \qst\, however, if $x$ is an m-atom one cannot deduce that $x$ is precisely the only element of $\llbracket x \rrbracket_w$, for identity must be used for that. So, we can reason within \qst\ that we may have in $\llbracket x \rrbracket_w$ \ita{an m-atom} having determinate properties, without being able to specify which particular one it is.

The following theorem expresses the invariance by permutations in \qst, a result we shall employ below\footnote{Recall the motivations for our development of quantum mechanics inside \qst: we want to use the features of non-individuals that are encapsulated in quasi-set theory to somehow obtain quantum theory}:

\begin{thm}[Unobservability of Permutations] Let $x$ be a finite q-set such that
$\neg(x = [z]_t)$ for some $t$ and let $z$ be an m-atom such that $z \in x$. If $w \in t$, $w \equiv z$ and $w \notin x$, then there exists $\llbracket w \rrbracket _t$ such that
$$(x - \llbracket z \rrbracket_t) \cup \llbracket w \rrbracket_t \equiv x$$
\end{thm}\label{unobser}

In words: two indiscernible elements $z$ and $w$, with $z \in x$ and $w \notin x$, expressed by their strong-singletons $\llbracket t \rrbracket_t$ and $\llbracket w \rrbracket_t$, are `permuted' and the resulting q-set remains indiscernible from the original one. The hypothesis that $\neg(x = [z]_t)$ grants that there are in $t$ indiscernible elements from $z$ which do not belong to $x$ (for details and the proof of the Unobservability of Permutations theorem, see \cite[pp.295,296]{frekra06}).

These are the central notions that will be important in our development of \qst-spaces, to which we now turn below.

\subsection{Quasi-functions}

From now on, we shall follow \cite{domholkra08}, where further details can be seen.
We begin with a q-set of real numbers $\epsilon = \{\epsilon_{i}\}_{i \in I }$, where $I$ is an arbitrary collection of indexes, denumerable or not. Since it is a collection of real numbers, which may be constructed in the classical part of \qst\, we have that $Z(\epsilon)$. Intuitively, the elements $\epsilon_{i}$ represent the eigenvalues of a physical observable $\hat{O}$, that is, they are the values such that $\hat{O}|\varphi_{i}\rangle=\epsilon_{i}|\varphi_{i}\rangle$, with $|\varphi_{i}\rangle$ the corresponding eigenstates. The fact that the observables are Hermitian operators grants us that the eigenvalues are real numbers, thus we are justified in assuming $\epsilon$ to be a set of real numbers. Consider then the quasi-functions $f:\epsilon \longrightarrow \mathcal{F}_{p}$, where $\mathcal{F}_{p}$ is the quasi-set formed of all finite and pure quasi-sets (that is, finite quasi-sets whose only elements are indistinguishable m-atoms). Each of these $f$ is a q-set of ordered pairs $\langle \epsilon_{i}, x\rangle$ with $\epsilon_{i}\in\epsilon$ and $x\in\mathcal{F}_{p}$. From $\mathcal{F}_{p}$ we select those quasi-functions $f$ which attribute a non-empty q-set only to a finite number of elements of $\epsilon$, the image of $f$ being $\emptyset$ for the other cases. We call $\mathcal{F}$ the quasi-set containing only these quasi-functions. Then, the quasi-cardinal of most of the q-sets attributed to elements of $\epsilon$ according to these quasi-functions is $0$. Now, elements of $\mcal{F}$ are quasi-functions which we read as attributing to each $\epsilon_{i}$ a q-set whose quasi-cardinal we take to be the occupation number of this eigenvalue. We write these quasi-functions as $f_{\epsilon_{i_{1}}\epsilon_{i_{2}}\ldots\epsilon_{i_{m}}}$. According to the given  intuitive interpretation, the levels $\epsilon_{i_{1}}\epsilon_{i_{2}}\ldots\epsilon_{i_{m}}$ are occupied. We say that if the symbol $\epsilon_{i_{k}}$ appears $j$-times, then the level $\epsilon_{i_{k}}$ has occupation number $j$. For example, the notation $f_{\epsilon_{1}\epsilon_{1}\epsilon_{1}\epsilon_{2}\epsilon_{3}}$ means that the level $\epsilon_{1}$ has occupation number $3$ while the levels $\epsilon_{2}$ and $\epsilon_{3}$ have occupation numbers $1$. The levels that do not appear have occupation number zero. Another point to be remarked is that since the elements of $\epsilon$ are real numbers, we can take the standard ordering relation over the reals and order the indexes according to this ordering in the representation $f_{\epsilon_{i_{1}}\epsilon_{i_{2}}\ldots\epsilon_{i_{m}}}$. \label{or} This will be important when we consider the cases for bosons and fermions.

The quasi-functions of $\mcal{F}$ provide the key to the solution to the problem of labeling states. In fact, since we use pure quasi-sets as the images of the quasi-functions, there is simply no question of indexes for particles, for all that matters are the quasi-cardinals representing the occupation numbers. To make it clear that permutations change nothing, one needs only to notice that a quasi-function is a q-set of weak ordered pairs (see definitions \ref{wp} and \ref{qf}). Taking two of the pairs belonging to some quasi-function, let us say $\langle\epsilon_{i}, x \rangle$, $\langle \epsilon_{j}, y\rangle$, with both $x$ and $y$ non-empty, a permutation of particles would consist in changing elements from $x$ with elements from $y$. But, by the unobservability of permutations theorem \ref{unobser}, what we obtain after the permutation is a q-set indistinguishable from the one we began with. Remember also that a quasi-function attributes indistinguishable images to indistinguishable items; thus, the indistinguishable q-set resulting from the permutations will also be in the image of the same eigenvalue. To show this point precisely, we recall that by definition $\langle\epsilon_{i}, x\rangle$ abbreviates $[[\epsilon_{i}],[\epsilon_{i},x]]$,\footnote{We are leaving aside the subindices in this notation.} and an analogous expression holds for $\langle\epsilon_{j}, y\rangle$. Also by definition, $[\epsilon_{i},x]$ is the collection of all the items indistinguishable from $\epsilon_{i}$ or from $x$ (taken from a previously given q-set). For this reason, if we permute $x$ with $x'$, with $x\equiv x'$ we change nothing for $[\epsilon_{i},x] \equiv [\epsilon_{i},x']$. Thus, we obtain $\langle\epsilon_{i},x\rangle \equiv \langle\epsilon_{i},x'\rangle$ and the ordered pairs of the `permuted' quasi-function will be indiscernible (the same if there are no m-atoms involved). Thus, the permutation of indistinguishable elements does not produce changes in the quasi-functions.

\subsection{A Vector Space Structure}

Now, we wish to have a vector space structure to represent quantum states. To do that, we need to define addition and multiplication by scalars. Before we go on, we must notice that we cannot define these operations directly on the q-set $\mcal{F}$, for there is no simple way to endow it with the required structure; our strategy here is to define $\star$ (multiplication by scalars) and $+$ (addition of vectors) in a q-set whose vectors will be quasi-functions from $\mcal{F}$ to the set of complex numbers $\mathbb{C}$. Let us call $C$ the collection of quasi-functions which assign to every $f\in \mathcal{F}$ a complex number. Once again, we select from $C$ the sub-collection $C_{F}$ of quasi-functions $c$ such that every $c\in C_{F}$ attributes complex numbers $\lambda \neq 0$ for only a finite number of $f\in \mathcal{F}$. Over $C_{F}$ we can define a sum and a product by scalars in the same way as it is usually done with functions as follows.

\begin{dfn}
Let $\gamma$ $\in \mcal{C}$, and $c$, $c_{1}$ and $c_{2}$ be  quasi-functions of $C_{F}$, then
$$(\gamma\star c)(f) \igual \gamma(c(f))$$
$$(c_{1}+ c_{2})(f) \igual  c_{1}(f) + c_{2}(f)$$
\end{dfn}

The quasi-function $c_{0}\in C_{F}$ such that $c_{0}(f)=
0$ for every $f\in \mcal{F}$ acts as the null element for the sum operation. This can be shown as follows:

\begin{equation}
(c_{0}+c)(f)= c_{0}(f)+c(f)= 0+c(f)= c(f), \forall f.
\end{equation}

With both the operations of sum and multiplication by scalars defined as above we have that $\langle C_{F},\mathbb{C}, +,\star \rangle$ has the structure of a complex vector space, as one can easily check. Some of the elements of $C_F$ have a special status though; if $c_{j}\in C_{F}$ are the quasi-functions such that $c_{j}(f_{i})= \delta_{ij}$ (where $\delta_{ij}$ is the Kronecker symbol), then the vectors $c_{j}$ are called the basis vectors, while the others are linear combinations  of them. For notational convenience, we can introduce a new notation for the q-functions in $C_{F}$; suppose $c$ attributes a $\lambda \not= 0$ to some $f$, and $0$ to every other quasi-function in $\mcal{F}$. Then, we propose to denote $c$ by $\lambda f$. The basis quasi-functions will be denoted simply $f_{i}$, as one can check. Now, multiplication by scalar $\alpha$ of one of these quasi-functions, say $\lambda f_{i}$ can be read simply as $(\alpha \cdot \lambda) f_{i}$, and sum of quasi-functions $\lambda f_{i}$ and $\alpha f_{i}$ can be read as $(\alpha + \lambda) f_{i}$. What about the other quasi-functions in $C_{F}$? We can extend this idea to them too, but with some care: if, for example $c_{0}$ is a quasi-function such that $c_{0}(f_{i})= \alpha$ and $c_{0}(f_{j})= \lambda$, attributing $0$ to every other quasi-function in $\mcal{F}$, then $c_{0}$ can be seen as a linear combination of quasi-functions of a basis; in fact, consider the basis quasi-functions $f_{i}$ and $f_{j}$, (this is an abuse of notation, for they are representing quasi-functions in $C_F$ that attribute 1 to each of these quasi-functions). The first step consists in multiplying them by $\alpha$ and $\lambda$, respectively, obtaining $\alpha f_{i}$ and $\lambda f_{j}$ (once again, this is an abuse, for these are quasi-functions in $C_F$ that attribute the mentioned complex numbers to $f_{i}$ and to $f_{j}$). Now, $c_{0}$ is in fact the sum of these quasi-functions, that is, $c_{0} = \alpha f_{i} + \lambda f_{j}$, for this is the function which does exactly what $c_{0}$ does. One can then extend this to all the other quasi-functions in $C_F$ as well.

\subsection{Inner Products}

The next step in our construction is to endow our vector space with an inner product. This is a necessary step for we wish to calculate probabilities and mean values. Following the idea proposed in \cite{domholkra08}, we introduce two kinds of inner products, which lead us to two Hilbert spaces, one for bosons and another for fermions. We begin with the case for bosons:

\begin{dfn}
Let $\delta_{ij}$ be the Kronecker symbol and
$f_{\epsilon_{i_{1}}\epsilon_{i_{2}}\ldots\epsilon_{i_{n}}}$ and
$f_{\epsilon_{i'_{1}}\epsilon_{i'_{2}}\ldots\epsilon_{i'_{m}}}$ two
basis vectors (as discussed above), then
\begin{equation}
f_{\epsilon_{i_{1}}\epsilon_{i_{2}}\ldots\epsilon_{i_{n}}}\circ
f_{\epsilon_{i'_{1}}\epsilon_{i'_{2}}\ldots\epsilon_{i'_{m}}} \igual
\delta_{nm}\sum_{p}\delta_{i_{1}pi'_{1}}\delta_{i_{2}pi'_{2}}\ldots\delta_{i_{n}pi'_{n}}.
\end{equation}
\end{dfn}

Notice that this sum is extended over all the permutations of the index set
$i'=(i'_{1},i'_{2},\ldots,i'_{n})$; for each permutation $p$,
$pi'=(pi'_{1},pi'_{2},\ldots,pi'_{n})$.

For the other vectors, the ones that can be seen as linear combinations in the sense discussed above, we have:
\begin{equation}
(\sum_{k}\alpha_{k}f_{k})\circ(\sum_{k}\alpha'_{k}f'_{k}) \igual
\sum_{kj}\alpha_{k}^{\ast}\alpha'_{j}(f_{k}\circ f'_{j}),
\end{equation}

\noindent where $\alpha^{\ast}$ is the complex conjugate of $\alpha$. Now, let us consider fermions. As remarked above in page \pageref{or}, the order of the indexes in each $f_{\epsilon_{i_{1}}\epsilon_{i_{2}}\ldots\epsilon_{i_{n}}}$ is determined by the canonical ordering in the real numbers. Thus, we define another ``$\bullet$'' inner product as follows, which will do the job for fermions:

\begin{dfn}
Let $\delta_{ij}$ be the Kronecker symbol and
$f_{\epsilon_{i_{1}}\epsilon_{i_{2}}\ldots\epsilon_{i_{n}}}$ and
$f_{\epsilon_{i'_{1}}\epsilon_{i'_{2}}\ldots\epsilon_{i'_{m}}}$ two basis vectors, then
\begin{equation}
f_{\epsilon_{i_{1}}\epsilon_{i_{2}}\ldots\epsilon_{i_{n}}}\bullet
f_{\epsilon_{i'_{1}}\epsilon_{i'_{2}}\ldots\epsilon_{i'_{m}}} \igual
\delta_{nm}\sum_{p}\sigma_{p}\delta_{i_{1}pi'_{1}}\delta_{i_{2}pi'_{2}}\ldots\delta_{i_{n}pi'_{n}}
\end{equation}
where:
$\sigma_{p} = 1$ if p is even and $\sigma_{p} = -1$ if p is odd.
\end{dfn}

This definition can be extended to linear combinations as in the previous case.

\subsection{Fock spaces using \qst-spaces}

We begin with a definition to simplify the notation. For every function $f_{\epsilon_{i_{1}}\epsilon_{i_{2}}\ldots\epsilon_{i_{n}}}$ in $\mcal{F}$, we put:

$$\alpha |\epsilon_{i_{1}}\epsilon_{i_{2}}\ldots\epsilon_{i_{n}}) \igual \alpha f_{\epsilon_{i_{1}}\epsilon_{i_{2}}\ldots\epsilon_{i_{n}}}$$

Note that this is a slight modified version of the standard notation. We begin with the case of bosons.

Suppose a normalized vector $|\alpha \beta \gamma \ldots )$, where the norm is taken from the corresponding inner product. Let $\zeta$ stand for an arbitrary collection of indexes. We define $a_{\alpha}^{\dagger}|\zeta ) \propto |\alpha \zeta )$ in such a way that the proportionality constant satisfies $a_{\alpha}^{\dagger} a_{\alpha} |\zeta ) = n_{\alpha} |\zeta )$. From this it will follow, as usual, that: $$((\zeta|a_{\alpha}^{\dagger})(a_{\alpha}|\zeta)) = n_{\alpha}.$$

\begin{dfn}
$a_{\alpha} | \ldots n_{\alpha} \ldots ) := \sqrt{n_{\alpha}} | \ldots n_{\alpha} - 1 \ldots )$
\end{dfn}

On the other hand $$a_{\alpha}a_{\alpha}^{\dagger} | \ldots n_{\alpha} \ldots ) = K \sqrt{n_{\alpha} + 1}|\ldots n_{\alpha} \ldots ),$$ where $K$ is a proportionality constant. Applying $a_{\alpha}^{\dagger}$ again, we have $$a_{\alpha}^{\dagger}a_{\alpha}a_{\alpha}^{\dagger} | \ldots n_{\alpha} \ldots ) = K^2 \sqrt{n_{\alpha} + 1}|\ldots n_{\alpha} + 1 \ldots ).$$

Using the fact that $a_{\alpha}^{\dagger} a_{\alpha} |\zeta ) = n_{\alpha} |\zeta )$, we have that: $$(a_{\alpha}^{\dagger}a_{\alpha})a_{\alpha}^{\dagger} | \ldots n_{\alpha} \ldots ) = \sqrt{n_{\alpha} + 1}K|\ldots n_{\alpha} + 1 \ldots ).$$ So, $K = \sqrt{n_{\alpha} + 1}$.

Then, we have:

\begin{dfn}
$a_{\alpha}^{\dagger} | \ldots n_{\alpha} \ldots ) :=  \sqrt{n_{\alpha} + 1}| \ldots n_{\alpha} + 1 \ldots ).$
\end{dfn}

From this definition, with additional computations, we obtain: $(a_{\alpha}a_{\beta}^{\dagger} - a_{\beta}^{\dagger}a_{\alpha})|\psi) = \delta_{\alpha \beta}|\psi)$. In our language, this means the same as $$[a_{\alpha} ; a_{\beta}^{\dagger}] = \delta_{\alpha \beta} I.$$

In an analogous way, it can be shown that:$$[a_{\alpha}; a_{\beta}] = [a_{\alpha}^{\dagger};a_{\beta}^{\dagger}] = 0.$$

So, the bosonic commutation relation are the same as in standard Fock space formalism.

For fermionic states we use the antisymmetric product $\bullet$. We begin by defining the creation operator $C_{\alpha}^{\dagger}$:

\begin{dfn}
If $\zeta$ is a collection of indexes of non-null occupation numbers, then $C_{\alpha}^{\dagger} := \alpha |\zeta)$
\end{dfn}

If $\alpha$ is in $\zeta$, then $|\alpha \zeta)$ is a vector of null norm. This implies that $(\psi|\alpha \zeta) = 0$, for every $\psi$. It follows that systems in states of null norm have no probability of being observed. Furthermore, their addition to another vector does not contribute with any observable difference. To take the situation into account, we define:

\begin{dfn}
Two vectors $|\phi)$ and $|\psi)$ are similar if the difference between them is a linear combination of null norm vectors. We denote similarity of $|\phi)$ and $|\psi)$ by $|\phi) \cong |\psi)$.
\end{dfn}

Using the definition of $C_{\alpha}^{\dagger}$ we can describe what is the effect of $C_{\alpha}$ over vectors: $$(\zeta| C_{\alpha} := (\alpha \zeta|.$$

Then, for any vector $|\psi)$, $$(\zeta| C_{\alpha}|\psi) = (\alpha \zeta|\psi) = 0$$ for $\alpha \in \zeta$ or $(\psi|\alpha \zeta) = 0$. Then, if $|\psi) = |0)$, then $(\zeta| C_{\alpha}|0) = (\alpha \zeta|0) = 0$. So, $C_{\alpha}|0)$ is orthogonal to any vector that contais $\alpha$, and also to any vector that does not contain $\alpha$, so that it is a linear combination of null norm vectors. So, we can put by definition that $\vec{0} := C_{\alpha}|0)$. In an analogous way, if $ \sim \alpha$ denotes that $\alpha$ has occupation number zero, then we can also write $C_{\alpha}|(\sim \alpha) \ldots) = \vec{0}$, where the dots mean that other levels have arbitrary occupation numbers.

Now, using our notion of similar vectors, we can write $C_{\alpha}|0) \cong \vec{0}$ and $C_{\alpha}|(\sim \alpha) \ldots) \cong \vec{0}$. The same results are obtained when we use $\cong$ and the sign of identity. By making $|\psi) = |\alpha)$, we have $(\zeta| C_{\alpha}|\alpha) = (\alpha \zeta|\alpha) = 0$ in every case, except when $|\zeta) = |0)$. In that case, $(0| C_{\alpha}|\alpha) = 1$. Then, it follows that $C_{\alpha}|\alpha) \cong 0$. In an analogous way, we obtain $C_{\alpha}|\alpha \zeta) = \cong |(\sim \alpha) \zeta)$ when $\alpha \notin \zeta$. In the case $\alpha \in \zeta$, $|\alpha \zeta)$ has null norm, and so, for every $|\psi)$: $$(\alpha \zeta| C_{\alpha}^{\dagger}|\psi) = (\alpha \zeta|\alpha \psi) = 0.$$

It then follows that $$(\psi|C_{\alpha}|\alpha \zeta) = 0,$$ so that $C_{\alpha}|\alpha \zeta)$ has null norm too.

Now we calculate the anti-commutation relation obeyed by the fermionic creation and annihilation operators. We begin calculating the commutation relation between $C_{\alpha}$ and $C_{\beta}^{\dagger}$. We do that by studying the relationship between $|\alpha \beta)$ and $|\beta \alpha)$. Let us consider the sum $|\alpha \beta) + |\beta \alpha)$. The product of this sum with any vector distinct from $|\alpha \beta)$ is null. For the product with $|\alpha \beta)$ we obtain $(\alpha \beta|[|\alpha \beta) + |\beta \alpha)] = (\alpha \beta || \alpha \beta) + (\alpha \beta || \beta \alpha)$. By definition, this is equal to $\delta_{\alpha \alpha} \delta_{\beta \beta} - \delta_{\alpha \beta}\delta_{\beta\alpha} + \delta_{\alpha \beta}\delta_{\alpha\alpha} - \delta_{\alpha \alpha}\delta_{\beta\beta}$. This is equal to $1 - 0 + 0 - 1 = 0$.

The same conclusion holds if we multiply the sum $|\alpha \beta) + |\beta \alpha)$ by $(\beta \alpha|$. It then follows that $|\alpha \beta) + |\beta \alpha)$ is a linear combination of null norm vectors, which we denote by $|nn)$, so that $$|\alpha \beta) = - |\beta \alpha) + |nn).$$

Given that, we can calculate $$C_{\alpha}^{\dagger}C_{\beta}^{\dagger}|\psi) = |\alpha \beta\psi) = -|\beta \alpha |\psi) + |nn) = - C_{\beta}^{\dagger}C_{\alpha}^{\dagger}|\psi) + |nn).$$

From this it follows that $\{C_{\alpha}^{\dagger};C_{\beta}^{\dagger}\}|\psi) = |nn)$. We do not lose generality by setting $\{C_{\alpha}^{\dagger};C_{\beta}^{\dagger}\}|\psi) = 0 $. In an analogous way we conclude: $$\{C_{\alpha};C_{\beta}\}|\psi) = 0.$$

Now we calculate the commutation relation between $C_{\alpha}$ and $C_{\beta}^{\dagger}$. There are some cases to be considered. We first assume that $\alpha \neq \beta$. If $\alpha \notin \psi$ or $\beta \in \psi$ then $$\{C_{\alpha};C_{\beta}^{\dagger}\}|\psi) \approx \vec{0}.$$

If $\alpha \in \psi$ and $\beta \notin \psi$, assuming that $\alpha$ is the first symbol in the list of $\psi$, then $\{C_{\alpha};C_{\beta}^{\dagger}\}|\psi) = C_{\alpha}|\beta \psi) + C_{\beta}^{\dagger}|\psi(\sim \alpha))\cong -|\beta \psi (\sim \alpha)) + |\beta \psi (\sim \alpha)) = \vec{0}$.

Now, if $\alpha = \beta$ and $\alpha \in \psi$, then $\{C_{\alpha};C_{\alpha}^{\dagger}\}|\psi) = C_{\alpha}|\alpha \psi) + C_{\alpha}^{\dagger}|\psi(\sim \alpha))\cong \vec{0} + |\psi) = |\psi)$.

If $\alpha = \beta$ and $\alpha \notin \psi$, then $\{C_{\alpha};C_{\alpha}^{\dagger}\}|\psi) = C_{\alpha}|\alpha \psi) + C_{\alpha}^{\dagger}|\psi(\sim \alpha))\cong |\psi) + \vec{0} = |\psi)$.

In any case, we recover $\{C_{\alpha};C_{\alpha}^{\dagger}\}|\psi) \cong \delta_{\alpha \beta} |\psi)$. So, we can put $$\{C_{\alpha};C_{\alpha}^{\dagger}\} = \delta_{\alpha \beta}.$$

It then follows that the commutation properties in \qst-spaces are the same as in traditional Fock spaces.

Using this formalism, we can adapt all the developments done in \cite[Chap.7]{mat67} and \cite[Chap.20]{mer70} for the number occupation formalism. But here, contrary to what happens in these books, no previous (even unconscious) assumptions about the individuality of quantum objects is taken into account.

\section{Foundational and metaphysical remarks}\label{dis}

Now that a non-standard version of quantum mechanics was sketched inside quasi-set theory, what are the main philosophical lessons that one can draw from this whole enterprise? Are we supposed to substitute traditional quantum mechanics by its nonreflexive version? The answer to the last question is an obvious NO, or not necessarily. What is illustrated by this construction is only that one can take a version of non-individuality seriously and, building from there, to develop the formalism of quantum mechanics anew, without labeling and individuality. We remark once again that in using standard mathematics (standard set theory), we need to assume that the represented objects are individuals at the start, and need to make some mathematical tricks in order to keep them with something that mimic non-individuality. Once indiscernible non-individuals are essential to quantum mechanics, our move illustrates  that one can assume non-individuality and  indistinguishability as central in a formalism and, along with some set theoretical tools compatible with such non-individuality (\ita{i.e.} quasi-set theory), derive a version of the formalism of the theory.

This kind of move illustrates how one may profitably approach the cooperative work between the study of foundations of science and metaphysics. Metaphysics may be seen as contributing positively to the foundations of a scientific theory like quantum mechanics, by suggesting that due to the main features of the theory, we are dealing with a new category of entities, the non-individuals. Metaphysicians are entitled to describe the main metaphysical features that entities qualified as non-individuals could have. By searching for the logic that adequately describes those entities we are led to the development of \qst-spaces. Quantum mechanics, on its own side, contributes to the development of metaphysics. By being our most successful theory, and by putting so much weight on indiscernible entities, it forces metaphysicians to develop an adequate notion of non-individuals and indiscernible entities compatible with the description provided by quantum mechanics.

What we aim now is to present how we understand this kind of collaboration between science and metaphysics in a more general picture. The following  may be seen as a rudimentary approach to how a productive relation between science and metaphysics may be developed which does not focus only on science (such as Ladyman and Ross do in \cite{lr07}) and also does not rely so heavily on what metaphysicians have already achieved (as it seems to be the case of French's approach in \cite[chap.3]{fre14}). We intend to present how a really collaborative effort to understand quantum metaphysics, for instance, may proceed. This is not achieved by adopting neither a `physics first' approach, and nor by providing a `metaphysics first' approach. Both physics and metaphysics must be cooperatively involved in the enterprise. The `dialectics' of such a cooperation may be roughly described as a case of a process aiming at a state of Reflective Equilibrium. Let us see briefly how.

We begin, as usual, with an actually available formulation of our scientific theory; in this case, quantum mechanics. In general there is an informal interpretation of the theory. Anyway, the mathematics of the theory may give us some clues as to how the entities being described behave. In general, we simply talk about the entities described by the theory by transferring our everyday categories of stable objects, properties, relations and things in general to the domain being described by the scientific theory. What our most successful theories in the beginning of the twentieth century have shown us is that this transference is not always justified. General relativity, for instance, demanded for a revision in our concepts of space and time, while quantum mechanics demanded a revision of many other concepts as well, such as stable objects possessing properties and having well defined identity conditions (it is enough to recall some no-go theorems like Kochen-Specker's  and Bell's). That such is the case is, as is well-known, the main contention of Ladyman and Ross in \cite[Chap.1]{lr07}: metaphysicians have not yet taken seriously the lessons from modern physics.

However, after perceiving that a scientific theory demands a revision in some fundamental concepts, what are metaphysicians supposed to do? In the case of quantum mechanics, in the specific case of non-individuals, how can we rigorously describe the situation? Well, we must first recognize that the mathematical apparatus employed to describe the theory encapsulates some of the items requiring revision, like labeling and identity. So, how should we proceed? Just as suggested by French \cite{fre14} with his \ita{viking approach to metaphysics}, we may check what the metaphysicians can say about the new kind of entities. Have they already articulated a theory about the behavior of those entities? If they have, do those entities, thus categorized, mesh well with what the scientific theory says about the world? In the case of quantum entities, it seems, as we have argued in sections \ref{sta} and \ref{ind},  the metaphysical ideas of non-individuals, of indiscernible classes of entities, do not fit very well with some of the assumptions that are encoded in the standard formalism of the theory. So, in this case, the search for an alternative logical foundations is a first required step.

Our suggestion is that we could then use the information brought by the scientific theory to develop a new logic, a logic that already encodes the basic information about the entities dealt with. Nonreflexive logics do just that, as we have briefly sketched in the beginning of section \ref{qsp}. So, we use the actual formulation of a theory in order to discover what are the main features of the entities being dealt with. In case those features demand a revision of some of our most general categories, then we have at least two options: either (i) we reformulate the theory in order to keep the traditional categories as much as possible (as it happens for instance, in Bohmian mechanics) while still retaining empirical adequacy, or else (ii) we try to take those novelties seriously, at face value as it were, and look for the best place to implement them. Given that the idea of non-individuals, as we have been dealing with them, is a general category, related with identity, it seemed that the underlying logic was the most appropriate place to introduce that category. Also, that change of view proved to be very fruitful: we could not only account for the non-individuality of quantum particles, but also show how that non-individuality is able to generate a formalism of quantum mechanics anew! Another way to account for the novelties, still concerning this second case, is not to change the formalism or the logic, but to keep the mathematics of the theory intact and provide for an interpretation of the formalism that somehow accounts for the novelties of the theory. One of such options is now known as \ita{wave function realism}. The main idea behind this interpretation seems to be that the wave function of the mathematical formalism is the reality of which the theory is about. In this case, the very concept of ordinary space is to be completely redefined, along with the concept of three dimensional objects. Even though this is an interesting approach, we shall not discuss it in what follows, keeping our focus on the notion of non-individuality (for more on wave function realism, see the discussion in the collection by Alyssa Ney and David Albert \cite{aly13}, \cite{kraare14}).

So, by reformulating the formalism of Fock spaces in terms of \qst-spaces, we are doing much more than merely paying lip service the a revisionary form of metaphysics: we show, somehow, that the concept of a non-individual is fundamental in at least one understanding of quantum mechanics. This is different from, for instance, the claim that some kind of paraconsistent logic may be useful for our understanding of quantum superpositions (as advanced in \cite{cos13}; see also the discussion in \cite{are15}). In this case, one interprets a physical fact in logico-metaphysical terms. Our claim is that the nonreflexive foundations of quantum mechanics developed here go one step further: we do not merely take the claims of quantum theory and re-interpret them in a non-classical logic system, but we begin with the non-classical system and inside it we develop the scientific theory itself. So, by reading some of the main features of the entities dealt with by the theory, we are able to use now those very features as the cornerstones for a new formulation of the theory, one that takes those features as central. This move closes a kind of circle: from a standard formulation of the theory we obtain some metaphysical feature of the entities. By laboring over those features, we develop a metaphysical approach to them that is rigorous enough to be encapsulated by a system of logic. Finally, by employing that system of logic we are able to develop a formalism that is able to express the theory.

Perhaps we could describe the kind of procedure developed here with the help of an analogy with the situation in foundations of mathematics, as presented by Kunen \cite[p.191]{kun09}. In discussing the foundations of mathematics, Kunen claims that logic must be developed \ita{twice}: we begin with our intuitions and intuitive rules of inference and, based solely on them, develop an informal set theory. Obviously, some logic is required in order to develop a set theory, and in the case of standard set theories this logic is informal classical logic. Now, inside the set theory we have just developed, we can begin anew and develop, in a completely rigorous way, \emph{using set theoretical tools}, our very system of logic. Now, inside such a set theory can we make sure that a reasonable model theory, for instance, is available. So, logic must be built twice: first, as an informal system, capable of grounding the development of a set theory and  such that it does not presuppose set theoretical concepts in its formulation. Then, inside the just developed set theory, we can develop, as a set theoretical entity, logic itself. In the second case, the tools of set theory are clearly available, and rigorous research is possible (for a discussion of that move concerning nonreflexive logic, see \cite{arkra09} and \cite{are14}).

A similar situation happens in the metaphysical foundations of quantum mechanics we have advanced. We begin with our intuitive conceptual apparatus, good enough for our everyday working of prediction and development of the theory. Perhaps that apparatus comprises a minimal constructive logic (as suggested by da Costa \cite{cos08}, such a logic would have the minimum for our actual development of science). With that apparatus we can develop logic and the mathematics required for quantum mechanics. Now, as we have discussed, quantum mechanics seems to demand a revision of the apparatus we began with.\footnote{Again, see the discussion in \cite{kraare15} for further information on those issues.} Logic must be built again. Now, however, differently from Kunen's description of the situation in mathematics, a new system of logic must be developed, not inside a classical set theory, but totally new, from scratch, taking into account the lessons from science. This is what we have attempted to provide here in the case of non-individuals. Obviously we do not require that every field of knowledge take nonreflexive logic into account; no, our everyday activities are rooted on our everyday experiences, for which identity does not seem to be eliminable, at least by this time. That can be interpreted in two ways: either nonreflexive logics takes the place of classical logic everywhere, with the proviso that it is the `classical' part of \qst\ that is being used, or else one accepts a pluralism in logic, claiming that distinct fields of knowledge demand distinct systems of logic. That is a interesting debate, but we shall not face it now.

Suffice to say, for now, that the construction of \qst-spaces is only a beginning to the study of the non-reflexive foundations of quantum mechanics. Nonreflexive logics still present philosophical challenges and advance some interesting problems in the foundations of quantum mechanics. It is up to us to develop those systems and discover whether the system can face the challenges.

\end{document}